\documentclass[aip,rsi,reprint,superscriptaddress]{revtex4-1}
\usepackage{graphicx}
\usepackage{tikz}
\usepackage{siunitx}
\usepackage{overpic}
\usepackage{microtype}

\begin{document}

\title{Cold atoms in micromachined waveguides: a new platform for atom-photon interactions}

\author{E.~Da~Ros}
\author{N.~Cooper}
\email[]{nathan.cooper@nottingham.ac.uk}
\author{J.~Nute}
\author{L.~Hackermueller}
\affiliation{\mbox{School of Physics and Astronomy, University of Nottingham, Nottingham, NG7 2RD, United Kingdom}}

\date{\today}

\begin{abstract}

Hybrid quantum devices, incorporating both atoms and photons, can exploit the benefits of both to enable scalable architectures for quantum computing and quantum communication, as well as chip-scale sensors and single-photon sources. Production of such devices depends on the development of an interface between their atomic and photonic components. This should be compact, robust and compatible with existing technologies from both fields. Here we demonstrate such an interface. Cold cesium atoms are trapped inside a transverse, 30~$\mu$m diameter through-hole in an optical fiber, created via laser micromachining. When the guided light is on resonance with the cesium $D_2$ line, up to 87\% of it is absorbed by the atoms. The corresponding optical depth per unit length is $\sim700$~cm$^{-1}$, higher than any reported for a comparable system. This is important for miniaturization and scalability. The technique can be equally effective in optical waveguide chips and other existing photonic systems, providing a new platform for fundamental research.

\end{abstract}


\maketitle


\section{Introduction }
\label{Intro}
Interfacing cold atoms with optical waveguides has been a very active field of research over the past two decades \cite{Duan2001,Kimble2008,Lepert2011,Thompson2013,Lee2013,Ritter2018,Hilton2018}. This has been motivated by the great potential of hybrid atom-photon systems for sensing, quantum communication and quantum information processing~\cite{Kimble2008,Monro2001,Reiserer2015,Nieddu2016}. Existing approaches tend to involve purpose designed systems such as tapered nanofibres~\cite{Sague2007,Nayak2008,Kumar2015,Patterson2018,Corzo2019}, hollow-core fibres \cite{Renn1995,Bajcsy2009,Okaba2014,Blatt2016,Langbecker2018} or other custom-built waveguide and photonic crystal structures~\cite{Kohnen2010,Hood2016}, which don't lend themselves readily to integration. Herein we describe a universal technique, capable of interfacing cold atoms with nearly any existing waveguide system. This new approach will allow hybrid devices to take full advantage of the capabilities of all-optical waveguide chips, which have reached a very high level of development \cite{Marshall2009,Sansoni2010,Poulios2014}. The increased range and complexity of operations thus permitted can be expected to greatly expand the potential of hybrid atom-photon devices, opening the door to long-standing experimental and technological goals \cite{Duan2001,Kimble2008,Monro2001,Reiserer2015}.

\begin{figure*}[ht]
\includegraphics[width=1.8\columnwidth]{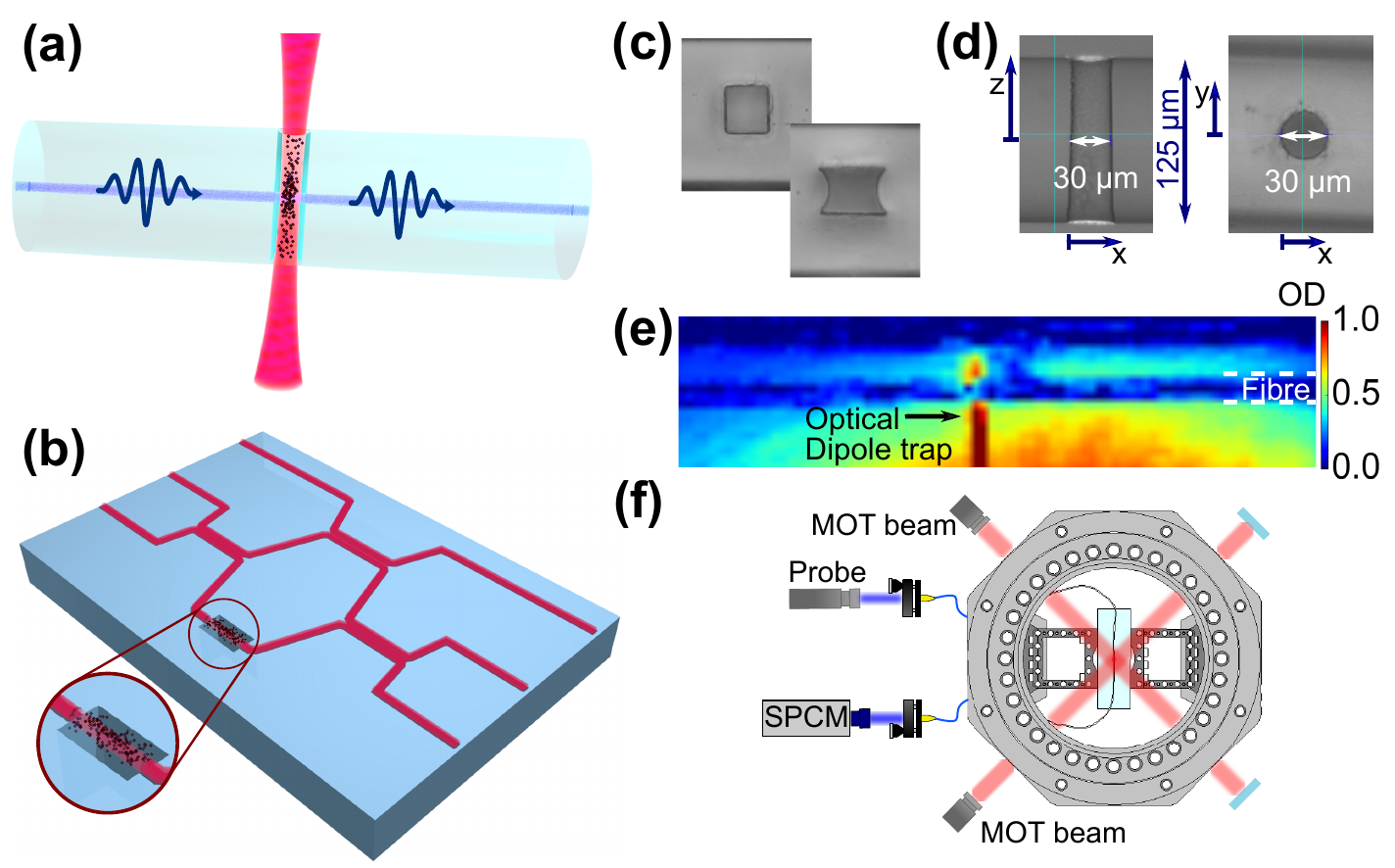}
\caption{\label{fig:project_aimim} \textbf{(a)}, Illustration of an atom cloud confined within a microscopic hole in an optical fiber using an optical dipole trap (red beam) and interfaced with light guided in the fiber (blue beam). \textbf{(b)}, Illustration showing the proposed use of this technique for the inclusion of cold atoms in a waveguide chip. \textbf{(c)} and \textbf{(d)}, Examples of differently shaped laser drilled holes and side and top view of the hole used in this study, images by Workshop of Photonics. \textbf{(e)}, Absorption image of cesium atoms confined in an optical dipole trap aligned with the void in the fiber (fiber visible as a horizontal feature at the center of the image). \textbf{f}, Schematic of the setup (top view). Atoms are trapped in a magneto-optical trap below the interface, then guided to the interaction region via an optical dipole trap. Here they are probed by light propagating in the fiber (blue beam). The output is collected by a single photon counting module. }
\end{figure*}

The basis of the technique described herein is the introduction of cold atoms into microscopic, laser-drilled holes in optical waveguides, see Fig.~\ref{fig:project_aimim}\textbf{(a)} and \ref{fig:project_aimim}\textbf({b)}. Laser micromachining can be used to create holes at any desired location(s) within a very wide range of materials, including most glasses and ceramics, silicon, metals and polymers \cite{Mishra2015}. This is a significant advantage over techniques linked to specific waveguide materials and architectures, as it will allow the method to be applied easily across a broad range of waveguide devices and to take full advantage of advances in subwavelength integrated optics and optical metamaterials \cite{Cheben2018}. Furthermore, the shape of these holes can be controlled and modified according to the needs of the user\cite{Cooper2019}, see Fig.~\ref{fig:project_aimim}\textbf{(c)}. 

Drilling a transverse hole through an optical waveguide allows cold atoms to be directly overlapped with the centre of the propagating mode of the waveguide and hence enables good coupling between photons and atoms. It has been shown that with the addition of an optical cavity strong coupling between the atoms and the guided light is achievable in such systems \cite{Cooper2019}. The technique can be extended to multiple holes in a single waveguide, thus allowing atoms at different sites to be coherently coupled via the optical field. 
 
We provide an experimental demonstration of the above technique within a standard, single-mode optical fiber (Thorlabs 780HP). A cloud of cold Cs atoms is guided into 30 $\mu$m diameter cylindrical through-hole, drilled orthogonally to the core of the fiber. The atoms absorb 87\% of the optical power of a resonant probe beam traveling through the fiber.

The small size of the interaction region, which is some tens of micrometres in length, is a clear advantage over existing methods, e.g. many fibre-based experiments involve interaction regions with lengths on the centimetre scale. We measure the optical depth per unit length of our trapped atom cloud at $\sim700$~cm$^{-1}$, greater than any value found in the literature for a comparable system \cite{Kaczmarek2015,Dawkins2011}. This has important benefits for device miniaturisation, enabling scalability of the number of waveguide-atom interfaces within a small device and improving spatial resolution in sensing applications \cite{Monro2001}. In the directions transverse to the propagation axis, the interface retains the small mode area of the waveguide. The tight confinement of the light while traversing the interaction region naturally helps to achieve strong atom-photon coupling \cite{Tey2008,Bajcsy2009}.

The technique permits large atom-surface separations. This diminishes the influence of decoherence and line-broadening effects caused by atom-surface interactions, enabling the use of charged or highly polarisable particles. Rydberg atoms are an important example of such polarisable particles, with significant applications for sensing and multi-qubit atomic gate operations \cite{Urban2009,Petrosyan2017}.  

Furthermore, the ability to hold the atoms within the interaction region for a prolonged period using an optical dipole trap allows sequential or time-separated operations.

Overall, the approach described herein results in a compact, robust interface that can be directly inserted wherever it is needed within an existing waveguide architecture. Our results represent an important advance towards the goal of combining cold atoms and optical waveguides into an integrated and scalable quantum device.

\section{Experimental realisation}
\label{setup}
Our experimental setup is illustrated in Fig.~\ref{fig:project_aimim}. The interface is based on a commercially available single mode optical fiber with a mode field diameter of (5.0 $\pm$ 0.5)~$\mu $m at 850~nm and a cladding diameter of 125~$\mu $m. The fiber has been laser drilled with a cylindrical through hole perpendicular to the light propagation axis, with diameter $D=30~\mu$m, as shown in Fig.~\ref{fig:project_aimim}\textbf{(d)}. Laser drilling was performed by Workshop of Photonics. 

For mounting purposes, the fibre is attached to a glass slide, which is then placed into a vacuum chamber. The fibre enters the vacuum system through custom built fibre-feedthroughs \cite{Abraham1998} and the region containing the hole is suspended in the center of the vacuum chamber.
Approximately $3 \times 10^7$ $\mathrm{^{133}Cs}$ atoms are loaded into a magneto-optical trap (MOT), about 1~mm below the fiber. To cool the cloud further, the trapping lasers are then detuned by 13~$\Gamma$ in an 18~ms optical molasses stage (where $\Gamma$ is the natural linewidth of the transition). During this time the atoms are also optically pumped to the $\lbrace6^2S_{1/2}, F=4\rbrace$ state.
At the same time the magnetic field gradient used to generate the MOT is ramped down and the magnetic field center is displaced upwards by increasing an offset field.
This transport technique allows us to achieve an atom density $n\approx 5 \times 10^{10}$ atoms/cm$^3$ immediately below the hole in the fiber. 

A fiber laser with a wavelength of 1064 nm and a maximum power of 25~W is used to create an optical dipole trap for the atoms. The trap consists of a vertical beam which is focused down to a waist of 13 $\mu$m and aligned through the hole in the optical fiber, see Fig.~\ref{fig:project_aimim}\textbf{(e)}. The focal point lies within the interaction region. 
After switch-off of the MOT beams, a hold-time of 3~ms allows the atoms to migrate along the dipole trap beam into the interaction region.

Pulses of probe light are sent through the fiber so that they interact with the atoms. For this we use light resonant with the $\vert g\rangle = \lbrace6^2S_{1/2}, F=4\rbrace \longrightarrow \vert e\rangle = \lbrace 6^2P_{3/2}, F=5\rbrace$ transition of $^{\mathrm{133}}$Cs. One microsecond before the probing pulse, the dipole trapping beam is switched off. The output of the optical fiber is directed onto a single photon counting module (SPCM), as shown in Fig.~\ref{fig:project_aimim}\textbf{(f)}, which is used to determine the number of transmitted photons. 

In order to measure the fraction of the probe light absorbed by the atom cloud, we count the number of photons received during a 15~$\mu$s probe beam pulse. We then repeat the measurement after a 100 ms delay, during which the atoms are dispersed. The absorption fraction is determined from the ratio of the two results. In order to avoid saturating the atomic transition, very low probe beam powers (on the order of pW) are necessary, and as a result photon counting statistics become our dominant source of experimental random error, see Appendix~\ref{app:Measurement}. We average multiple measurements (typically between 30 and 100) for any given set of experimental conditions, thus reducing the statistical uncertainty.

\section{Results}
\label{results}
Using a probe beam power $P_{\mathrm{pr}}$= 13~pW, we measured $(87\pm 2)$\% absorption caused by the atoms, corresponding to an optical depth (OD) of $2.1\pm0.2$, leading to an optical depth per unit length of 700~cm$^{-1}$. This implies the presence of $290\pm20$ Cs atoms in a volume of 600~$\mu$m$^3$ and corresponds to an average density on the order of $5 \times 10^{11}$~atoms/cm$^3$, assuming a uniform atomic distribution across the interaction region. These values therefore represent a lower limit on the atom number and peak density.

\begin{figure}
\includegraphics[width=0.45\textwidth]{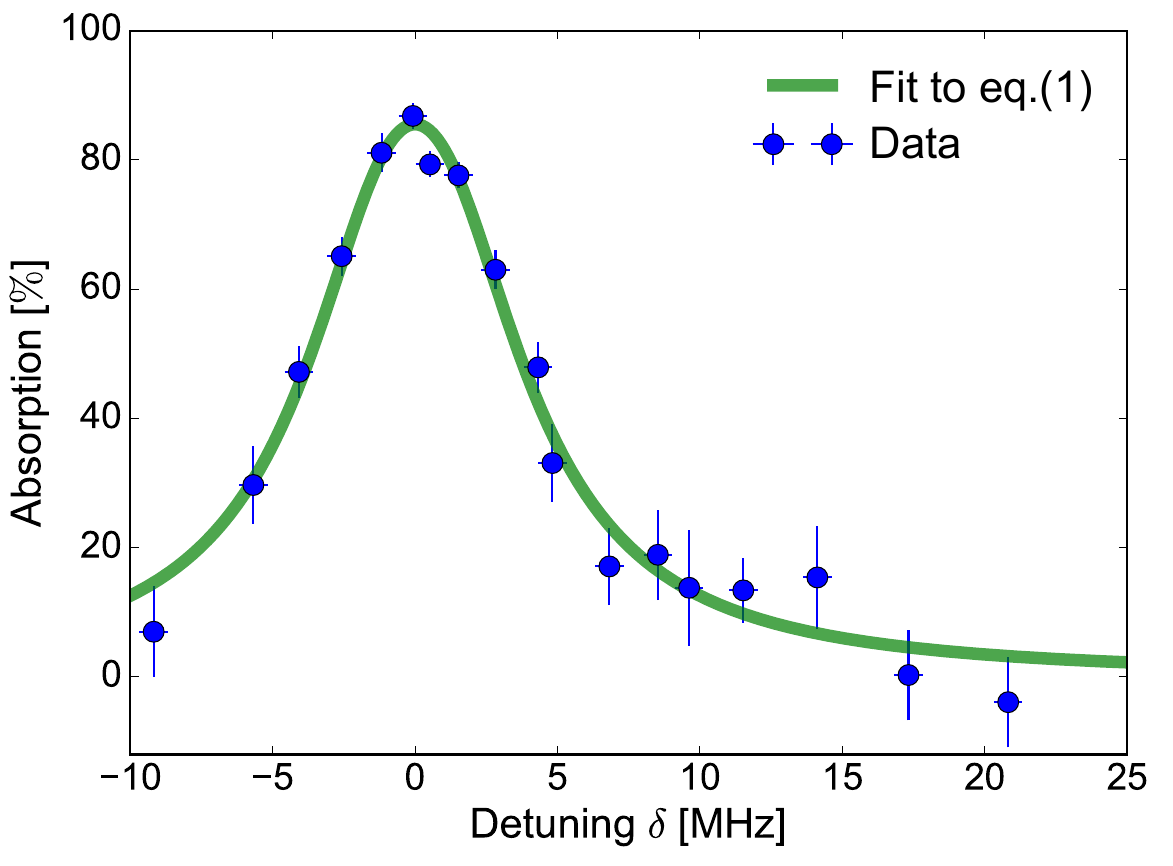}
\caption{\label{fig:Abs_vs_probe_detuning} Fractional absorption of the probe beam light by the atoms as a function of the detuning $\delta$ from the $\vert g\rangle = \lbrace6^2S_{1/2}, F=4\rbrace \longrightarrow \vert e\rangle = \lbrace 6^2P_{3/2}, F=5\rbrace$ $^{\mathrm{133}}$Cs transition. Each point corresponds to the mean of 36 repetitions, with the error bars representing the standard deviation. The solid line is a fit according to eq.~(\ref{eq:abs_detuning}).}
\end{figure}

The response of the atomic absorption to changes in probe laser detuning, probe laser power and dipole laser power was measured. Fig.~\ref{fig:Abs_vs_probe_detuning} shows the fractional absorption of the probe beam power by the atom cloud as a function of the probe laser detuning $\delta$, as the laser frequency is tuned across the $\lbrace F=4\rbrace \longrightarrow \lbrace F'=5\rbrace$ transition. The probe beam power was $P_{\mathrm{pr}}$= 9~pW. The resonance profile derives from the natural linewidth of the atomic transition $\Gamma$, with the atomic absorption response in percent given by \cite{Bajcsy2009}
\begin{equation}
\label{eq:abs_detuning} \mathrm{Abs}= 100 \left({ 1-\exp{\left( \frac{-\mathrm{OD}}{1+4(\frac{\delta}{\Gamma})^2}\right)} }\right).
\end{equation} 
This model is valid in the limit of low atomic saturation (i.e. the intensity of the probe light, $I$, must be very much lower than the saturation intensity, $I_{\mathrm{sat}}$ of the atomic transition), which is fulfilled in this case. 
The green line in Fig.~\ref{fig:Abs_vs_probe_detuning} represents a fit of eq.~(\ref{eq:abs_detuning}) to the experimental data, where natural linewidth and optical depth have been used as free parameters. This fit results in values of OD=~$1.9 \pm 0.2$ and $\Gamma$=~$2 \pi \times (5.5\pm0.4)$ MHz. The experimentally derived value for $\Gamma$ is in good agreement with the theoretical value of $2 \pi \times 5.2$~MHz, proving that the atomic transition is not noticeably broadened by the presence of the fiber. 

\begin{figure}
\includegraphics[width=0.45\textwidth]{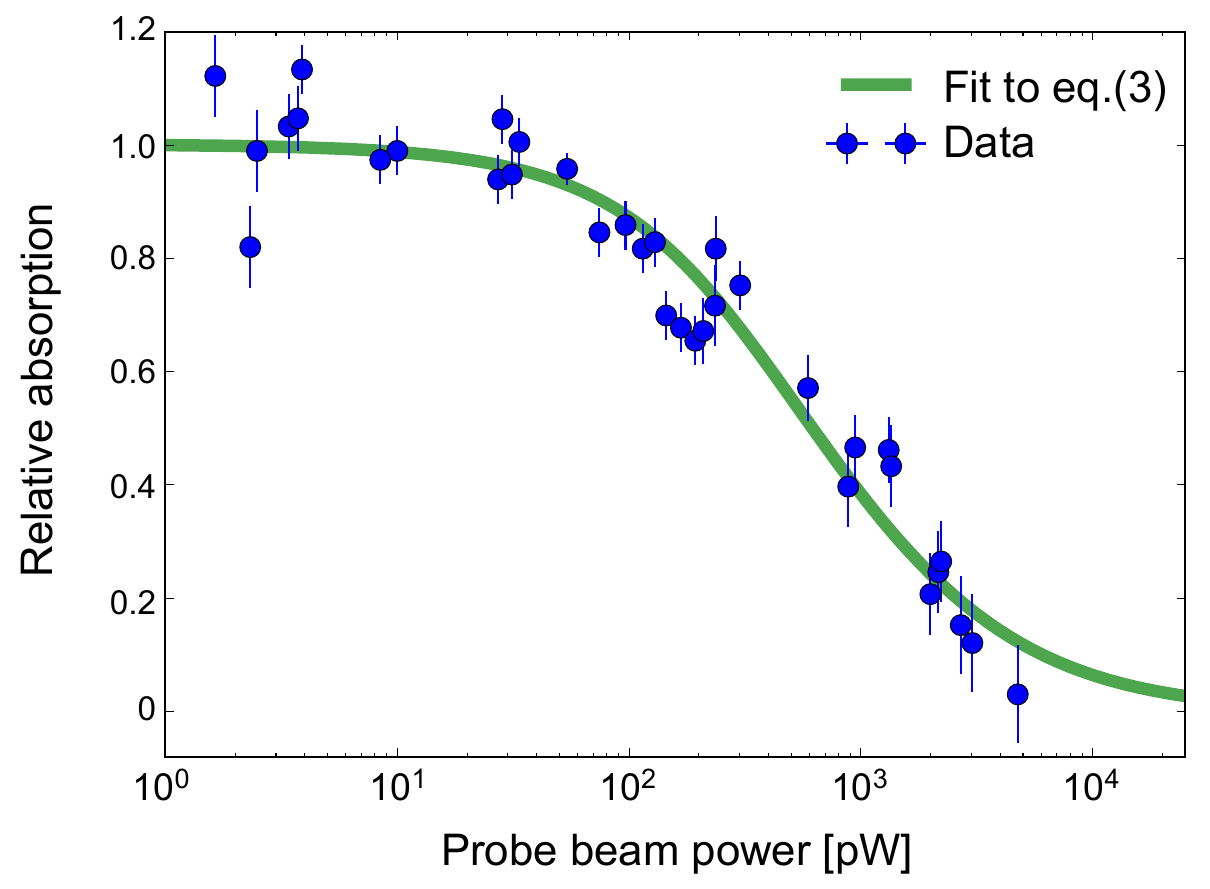}
\caption{\label{fig:Abs_vs_probe_power} Relative absorption of resonant probe light as a function of the incident probe beam power. Each point corresponds to the mean of 49 repetitions, with the error bars representing the standard deviation. The solid line is a fit according to eq.~(\ref{eq:abs_probe_power}).
}
\end{figure}

The dependence of the atomic absorption response on the probe beam power was characterized and the results are shown in Fig.~\ref{fig:Abs_vs_probe_power}. 
The fact that, in this case, neither the optical depth nor the ratio of the probe beam intensity to the atomic saturation intensity can be approximated as small necessitates a numerical approach to modeling this situation.

Consider an element of the incident intensity distribution at a specific position $I(0,y,z)$. Employing the standard assumption of direct ray propagation\cite{Goban2012,Vetsch2010}, the intensity $I$ of the light traversing the atom cloud will obey the differential equation 
\begin{equation}
\label{eq:diff_eq_sat}  
\frac{dI(x,y,z)}{dx}=-n \sigma_0\frac{I(x,y,z)}{1+\frac{I(x,y,z)}{I_{\mathrm{sat}}}}.
\end{equation} 

where $n$ represents the mean atomic density in the interaction region and $\sigma_0$ represents the atomic absorption cross section on resonance. The atomic density can be approximated as uniform along the $y$ and $z$ directions, see Appendix~\ref{app:Atom loading}. The distribution of the atoms along the $x$ axis, while non-uniform, does not affect the final light intensity resulting from our model. 
We solve eq.~(\ref{eq:diff_eq_sat}) numerically for $I(D,y,z)$, the probe light intensity transmitted through the hole at position $(y,z)$, where the incident optical intensity $I(0,y, z)$ is given by the input power $P$ and the mode profile of the fiber.
We assume that, in the $z$ axis as shown in Fig.~\ref{fig:project_aimim}\textbf{(d)}, the propagating light mode of the fiber is not significantly altered as it traverses the void. In the $y$ axis, the curvature of the hole acts as a cylindrical lens, leading to a divergence of the propagating beam mode. To account for this we use a simplistic model, where it is assumed that light incident too far from the fiber axis, at coordinates with $\vert y\vert$ greater than a cut-off distance $c_y$, will not be coupled back into the guided mode on the other side of the void. The cut-off distance $c_y$ is then used as a free parameter when fitting the data to the model. 

Integrating over the dimensions (y,z) then results in the total absorption:
\begin{equation}
\label{eq:abs_probe_power} \mathrm{Abs}= { 1-\frac{1}{P}\int\limits_{-c_z}^{c_z}\int\limits_{-c_y}^{c_y} I(D,y,z)~dy dz},
\end{equation} 
where $P$ is the total power within the integration area
\begin{equation}
\label{eq:probe_power} P=\int\limits_{-c_z}^{c_z}\int\limits_{-c_y}^{c_y} I(0,y,z)~dy dz,
\end{equation} 
$c_z$ is taken to be large compared to the mode radius of the fiber, and $I(0,y,z)$ takes into account the Gaussian distribution of the probe beam intensity. 

\begin{figure}
\includegraphics[width=0.45\textwidth]{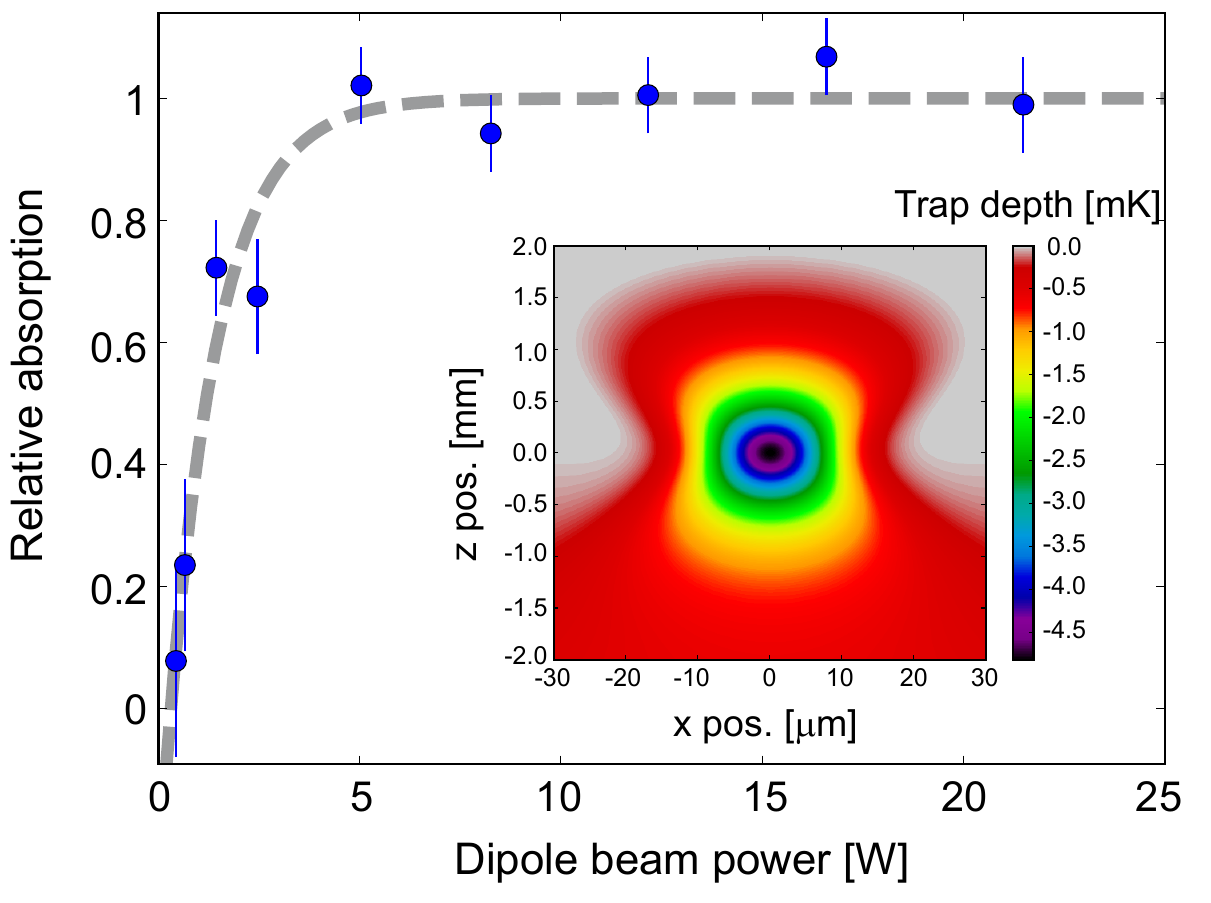}
\caption{\label{fig:Abs_vs_Nufern} Relative absorption of resonant probe light as a function of the power in the optical dipole trapping beam. Each point corresponds to the mean of 49 repetitions, with the error bars representing the standard deviation. The line plot is a guide to the eye. In the inset the trapping potential is depicted for 5\,W of trapping power. The asymmetry in the plot is due to gravity.}
\end{figure}

Fitting the data to this model, with the atom density $n$ and the $y$ cut-off distance $c_y$ as free parameters, gives the theoretical fit plotted in Fig.~\ref{fig:Abs_vs_probe_power}. The resulting value $c_y=$~(1.9 $\pm$ 0.3)~$\mu$m is comparable to the fibre mode radius, as expected. The corresponding atom density $n\approx $~(2.9~$\pm$~0.1)~$\times 10^{11}$~atoms/cm$^3$ implies the presence of 170~$\pm$~5~atoms in the interaction region. We assume that the light reaching the atoms is linearly polarized, and use the corresponding saturation intensity $I_{\mathrm{sat}}=$~1.66~mW/cm$^2$\hspace{0.15cm}, although small deviations from this are possible as the fiber is not polarization maintaining. 
Finally, the absorption of the probe beam was measured as a function of the power used in the dipole trapping beam, as shown in Fig.~\ref{fig:Abs_vs_Nufern}. The results show that, for powers in excess of 5~W (corresponding to trapping frequencies of $\nu_r~\approx 76$~KHz in the radial direction and $\nu_z~\approx 2.0$~KHz in the longitudinal direction), the number of atoms loaded into the interaction region is no longer limited by the power of the dipole trapping beam. This indicates that for powers larger than 5~W the depth of the optical dipole trap has become large compared to the temperature of the atomic cloud. The fact that the probe absorption appears to drop to zero for finite dipole beam powers is expected given the influence of gravity on the trapping potential.   

\section{Conclusions and outlook\label{Discussion}}
We have demonstrated efficient coupling of a cold atomic ensemble, introduced into a laser-micromachined hole in an optical waveguide, to the central mode of the light guided in the waveguide. This method for interfacing atoms and waveguides enables the establishment of new atom-photon interaction platforms. It is applicable within a wide range of waveguide architectures and offers high optical depth per unit length. This technique will have important applications in the production of miniaturized atom-optical devices for sensing, quantum communication and quantum information processing. 

Demonstration of this technique within a more complex waveguide network, together with its use to produce coherent atom-optical effects such as electromagnetically induced transparency \cite{Boller1991} or four-wave mixing \cite{Geng2014,Ding2015,Gulati2015}, represents the next step towards a quantum information processing device based on this architecture.

We have shown in \cite{Cooper2019} that appropriate shaping of the micromachined holes can lead to very high transmissions comparable to existing interfaces. New holes with tailored shapes leading to higher transmission efficiencies are going to be implemented in future designs of the interface. The reduced optical losses make conceivable to introduce optical cavities around such holes, for example via the use of laser written Bragg gratings in the waveguides \cite{Meltz1989,Kato2015}, with a sufficiently high finesse to bring the atom-light system into the strong coupling regime \cite{Kimble1998}. For atom numbers comparable to those demonstrated here, we find that cooperativities on the order of $C = 400$ should be possible \cite{Cooper2019}. This would enable photonically-mediated interactions between atoms located at distant sites, and also permits single photon gate operations. The technique thus has the potential to directly interface photonic and atomic qubits. 

\begin{acknowledgments}
\noindent The authors would like to thank E. Haller, K. Poulios, P. Verlot, G. Buonaiuto and the ErBeStA team for useful discussions. This project was supported by the EPSRC grants EP/R024111/1, EP/K023624/1, EP/M013294/1 and by the European Commission grants QuILMI (no. 295293) and ErBeStA (no. 800942). We acknowledge support from the University of Nottingham through a Birmingham-Nottingham collaboration grant.
\end{acknowledgments}

\section{Author contributions}
\noindent L.H. conceived the experiments and oversaw the work. System construction and testing was commenced by J.N. and N.C. and completed by N.C. and E.D.R.. E.D.R. conducted the experiment and collected the published data, with 
assistance from N.C.. E.D.R. and N.C. wrote the paper with assistance from L.H.. All authors contributed to the manuscript, E.D.R. and N.C. contributed equally to this work.

\appendix

\section{Interface details}
\label{app:Interface}
Prior to laser machining, a commercially available Thorlabs 780HP bare fiber is glued underneath a glass plate, using vacuum compatible UV-cure glue (Dymax Low-Shrink OP-67-LS). A 1 mm diameter, mechanically drilled hole in the glass plate allows unobstructed passage of the dipole and imaging beams through the interaction region. Following micromachining of the fiber, the chip is suspended in the center of the vacuum chamber, with the central region of the fiber free-hanging for approximately 5 mm.

In reference~\cite{Cooper2019} we describe how, with appropriate shaping, microscopic holes of the kind used herein can allow very high optical transmission efficiencies, comparable to similar platforms. For example, with a minimum separation of 30 micrometers between the two faces of the hole $>90$\% power transmission is possible, with the dominant source of loss being reflections at the fibre to hole interfaces. Our first prototype device employs a cylindrical hole and consequently offers a much lower transmission efficiency of $\sim20$\%. This results from a theoretical transmission efficiency of 31\% with the additional losses due to surface roughness inside the hole. Both of these problems are surmountable in future iterations, with micromachining of more favourable hole shapes having recently been accomplished (as shown in Fig.~\ref{fig:project_aimim}\textbf{(c)}) and transmissions close to the theoretical optimum for their chosen hole geometry having recently been achieved in silica-on-silicon waveguide platforms \cite{ErBeStA}.

The experiment has been used over the course of many months and no sign of degradation (i.e. broadening of the atomic linewidth due to coating of the hole's facets with Cs atoms) has been detected.

\section{Atom loading}
\label{app:Atom loading}
The MOT captures around $3 \times 10^7$ $\mathrm{^{133}Cs}$ atoms from the background gas within 10~s. The trap forms about 1~mm below the hole in the fiber. 

In order to cool the cloud further, the MOT beams are then detuned by $\sim$ 13~$\Gamma$ over 18~ms in a optical molasses stage. At the same time the current flowing in the two coils that generate the magneto-optical trap decreases with unbalanced ramps, modifying the field gradient in the region of the cloud. Simultaneously the magnetic offset field increases and the cloud is pushed upwards.
The presence of the glass plate and the scattering arising from the mechanically drilled hole in the mounting plate adversely affect the MOT beams and hence the MOT loading. The closer the starting position of the trap is to the surface of the chip, the lower the achievable atom density is. 
The transport stage is therefore necessary to obtain a dense, cold cloud close to the fiber.

The optical dipole trap is switched on 42~ms before the start of the optical molasses stage. The resulting 60~ms of overlap between the dipole trap and the MOT stage was found to maximize the dipole trap loading and the final atom number at the position of the fiber core.

The atoms are assumed to have a uniform distribution in the $y$ and $z$ directions. This is an reasonable assumption in our experiment, as both the Rayleigh length and the waist of the dipole trapping beam substantially exceed the diameter of the probe beam.

\section{Measurement methods}
\label{app:Measurement}
Between 1 and 10$^4$ pW of probe light is coupled into the optical fiber, and the fiber output is then focused onto the photoreceptor of a single photon counting module (Excelitas SPCM-AQRH-14). The number of photons reaching the SPCM is recorded using high frequency binary counter chips connected to an Arduino Uno.
A linearity correction function, specified by the manufacturer, is employed to account for the influence of the receptor dead time at high count rates.

In each experimental cycle, two readings are taken from the SPCM: the number of photons received with atoms present in the hole, $N_{at}$, and the number received without atoms, $N_{bgd}$. In both cases, the duration of the probe pulse is 15~$\mu$s, with 100~ms allowed between the two pulses for the atoms to disperse.

Unwanted SPCM counts arising from background light sources and scattered dipole trap light are minimized using a laser-line filter (SEMROCK LL01-852), positioned directly in front of the SPCM, and appropriate physical shielding. To compensate for the SPCM dark count rate and residual background light, a control data set is taken without any probe light. The mean number of counts recorded in this condition ($\sim$~1-2~counts) is then subtracted from all other measurements. 

The SPCM sensitivity can be modified following exposure to MOT light during the atom loading stage, and the entire experimental system is subject to small variations that occur repeatably over the course of an experimental cycle. To account for any systematic bias caused by these effects, the ratio of the counts recorded during the first and second probe pulses is also measured without atoms loaded into the hole, but with probe light in the fiber. To prevent atom loading, the magnetic field used to generate the MOT is switched off, but all other experimental parameters are unchanged. The ratio $N_{at}/N_{bgd}$ that is recorded with atomic loading is then divided by the mean ratio obtained in this condition, such that any remaining systematic difference between $N_{at}$ and $N_{bgd}$ must arise from the presence of the atoms.

The ratio of the two readings $N_{at}$ and $N_{bgd}$, after these corrections, identifies the transmission of the probe light through the atomic cloud. For statistical purposes each measurement is averaged over multiple repetitions, typically between 36 and 100.

The optical power incident on the Cs atoms in the hole, $P$, is estimated from the number of photons recorded by the SPCM without atoms in the hole, $N_{bgd}$. This estimate takes into account the quantum efficiency of the SPCM, the mode area of the fiber and the optical losses encountered on traversing the hole and subsequent components. It is assumed that loss of light at the hole occurs primarily due to limited coupling back into the guided mode.

%
\end{document}